
%
%
\def\half {{1 \over 2}}
\def\dz{{\partial _z}}
\def\dzbar{{\partial _{\bar z}}}

\def\va{{v^\alpha}}
\def\vb{{v^\beta}}
\def\vba{{\bar v^\alpha}}
\def\vbb{{\bar v^\beta}}
\def\m {{m^\mu}}
\def\plb{{+\bar l}}

\def\pmb{{+\bar m}}

\def\mm{{- m}}

\def\ml{{- l}}
\def\hp{{h^+}}
\def\hm{{h^-}}
\def\kp{{\kappa^+}}
\def\km{{\kappa^-}}
\def\kpm{{\kappa^\pm}}
\def\kmp{{\kappa^\mp}}
\def\Dp{{D_+}}
\def\Dm{{D_-}}
\def\Dpm{{D_\pm}}

\def\gmu {{\gamma^\mu_{\alpha\beta}}}
\def\lp {{\lambda^+}}
\def\lm {{\lambda^-}}
\def\lpm {{\lambda^\pm}}
\def\wp {{w^+}}
\def\wm {{w^-}}

\def\vepm {{\varepsilon^\pm}}

\def\hvep {{\hat\varepsilon^+}}
\def\hvem {{\hat\varepsilon^-}}
\def\hvepm {{\hat\varepsilon^\pm}}

\def\sp {{\psi^+}}
\def\sm {{\psi^-}}

\def\dzxp {{\dz x^{9+0} +\half\sp\dz\sm+\half\sm\dz\sp}}

\def\xplb {{x^\plb}}
\def\xml {{x^\ml}}
\def\Gplb {{\Gamma^\plb}}
\def\Gml {{\Gamma^\ml}}

\def\Gpmb {{\Gamma^\pmb}}

\def\Gmm {{\Gamma^\mm}}

\tolerance=5000
\footline={\ifnum\pageno>1
       \hfil {\rm \folio} \hfil
    \else \hfil \fi}

\overfullrule=0pt 
\baselineskip=18pt
\raggedbottom
\centerline{\bf Lorentz-Covariant Green-Schwarz Superstring Amplitudes}
\vskip 24pt
\centerline{Nathan Berkovits}
\vskip 24pt
\centerline{Math Dept., King's College, Strand, London, WC2R 2LS, United
Kingdom}
\vskip 12pt
\centerline{e-mail: udah101@oak.cc.kcl.ac.uk}
\vskip 12pt
\centerline {KCL-TH-92-6}
\vskip 12pt
\centerline {November 1992}
\vskip 96pt
\centerline {\bf Abstract}
\vskip 12pt

In a recent paper, the BRST formalism for the gauge-fixed
N=2 twistor-string was used to calculate Green-Schwarz supersring
scattering amplitudes with an arbitrary number of loops and
external massless states. Although the gauge-fixing procedure
preserved the worldsheet N=2 superconformal invariance of the
twistor-string, it broke the target-space SO(9,1) super-Poincar\'e
invariance down to an SU(4)xU(1) subgroup.

In this paper, generators for the SO(9,1) super-Poincar\'e
transformations, as well as manifestly covariant vertex operators,
are explicitly constructed out of the gauge-fixed matter fields.
The earlier calculated amplitudes are then expressed in manifestly
Lorentz-covariant notation.
\vfil
\eject
\centerline{\bf I. Introduction}
\vskip 12pt

There are various methods for calculating superstring amplitudes,
each having advantages and disadvantages. The most common method is
to start with the covariant Neveu-Schwarz-Ramond action in which all
gauge invariances have been fixed except for the worldsheet N=1
superconformal invariance. Since this action contains only free fields,
it is straightforward to calculate superstring scattering amplitudes
by defining vertex operators for the physical states, evaluating
correlation functions of these vertex operators on N=1 super-Riemann
surfaces, integrating over the super-moduli of the surfaces, and
finally summing over all possible spin structures.$^{1,2}$
These scattering amplitudes can be proven to be unitary by showing
agreement with amplitudes obtained using the light-cone NSR method.$^3$

One disadvantage of this approach is that until the final step of
summing over spin structures, the scattering amplitude is not
spacetime supersymmetric, and therefore contains divergences
coming from the dilaton tadpole diagram. In order to regularize
these divergences, a cutoff in the moduli space must be introduced
which can only be removed after summing over spin structures.
Because the integrand of the scattering amplitude changes by a
total derivative for different choices of anti-commuting moduli
for the surface,$^2$
this cutoff produces a boundary term that depends on the choice of
the anti-commuting moduli.$^4$ Although an unambiguous choice for the
anti-commuting moduli can be determined from unitary requirements,$^5$
the need to make such a choice complicates the analysis of the
multiloop scattering amplitudes.$^6$
A second disadvantage of the NSR approach is that the fermionic
vertex operator is a function of ghost fields, as well as matter fields.$^1$
This ghost dependence makes it awkward to formulate the NSR string
in a fermionic background, preventing a straightforward derivation of
the ten-dimensional supergravity equations of motion.

Another method for calculating superstring scattering amplitudes
is to start with the light-cone Green-Schwarz action in which all
gauge invariances including conformal invariance have been fixed.$^{7,8,9}$
This manifestly unitary
method differs in three important features from the covariant
Neveu-Schwarz-Ramond method. Firstly, there is no sum over spin
structures since spacetime supersymmetry is manifest. Secondly, the
action is defined on ordinary Riemann surfaces, rather than on
N=1 super-Riemann surfaces. And thirdly, non-trivial operators
need to be inserted at the interaction points of the surface in
order to preserve Lorentz invariance (these interaction points are
located at the zeros of $\dz\rho$, where $\dz\rho$ is the unique
meromorphic fuction with poles of residue $P^{9+0}_r$ at $z=z_r$
and purely imaginary periods when integrated around the $2g$
non-trivial cycles).

Although the first two features of the light-cone Green-Schwarz
method remove the problems of the covariant NSR method, the third
feature introduces new problems. Since correlation functions now
depend on the locations of the interaction points, as well as the
locations of the vertex operators, the resulting expressions for the
scattering amplitudes are
complicated functions of the moduli of the
surface. Furthermore, potential divergences when
two or more interaction points coalesce force the introduction of
contact terms,$^{9,10,11}$ whose contribution to multiloop Green-Schwarz
amplitudes$^9$ has not yet been included in any manifestly Lorentz-covariant
expressions.

It should be noted that the semi-light-cone Green-Schwarz method,$^{12}$
in which all gauge invariances are fixed except for conformal
invariance, has exactly the same advantages and disadvantages
as the light-cone Green-Schwarz method.$^{13}$ The only difference is that
in order to preserve Lorentz invariance, the non-trivial
operators need to be inserted at the zeros of the expectation value
of $\dz x^{9+0}$, since if one fixes the remaining conformal invariance
to get to light-cone gauge, $\dz x^{9+0}$ is replaced by $\dz\rho$.

Recently, a new method for calculating superstring amplitudes has
been developed in which one starts from an N=2 twistor-string
version of the Green-Schwarz action where all gauge invariances
have been fixed except for worldsheet N=2 superconformal invariance.$^{14,15}$
This method is manifestly spacetime supersymmetric like the light-cone
Green-Schwarz method, but requires no operator insertions at the
interaction points. Although the scattering amplitudes are calculated
by evaluating correlation functions on N=2 super-Riemann
surfaces, there is no ambiguity coming from the choice of anti-commuting
moduli since the dilaton tadpole diagram vanishes by spacetime
supersymmetry,\footnote
\dag{ The spacetime supersymmetry generators in the twistor-string
formalism
contain no ghost contributions, so the dilaton tadpole diagram
can be shown to vanish by writing the zero-momentum dilaton
vertex operator as the contour integral of a spacetime supersymmetry
generator around a zero-momentum dilatino vertex operator, and
pulling the contour off the back of the surface. In the
Neveu-Schwarz-Ramond formalism, the ghost contributions to the
spacetime supersymmetry generators have unwanted poles,$^2$ so the
same argument$^{16}$ implies only that the dilaton tadpole diagram is
a total derivative in moduli space.$^4$} and
there is therefore no need to introduce a cutoff
in the moduli space. Using standard BRST techniques, Green-Schwarz
superstring
scattering amplitudes with an arbitrary number of loops and
external states can be explicitly evaluated. These amplitudes agree
with those obtained using the manifestly unitary light-cone Green-Schwarz
formalism if one makes a simple conjecture concerning the contribution
of the light-cone gauge contact terms (although a similar conjecture
in the NSR formalism can be proven to be true, a proof in the
Green-Schwarz formalism is still lacking).

The only disadvantage of this new method is that in order to express
the N=2 twistor-string action in terms of free fields, the
manifest SO(9,1) super-Poincar\'e invariance has to be broken down
to an SU(4)xU(1) subgroup. However, it will be shown in this paper
that the twistor-string free fields can be combined into N=2 chiral and
anti-chiral superfields, $X^\mu_\pm$ and $\Psi^\alpha_\pm$, that
transform covariantly as SO(9,1) vectors and spinors under commutation
with the SO(9,1) Lorentz generators, $M^{\mu\nu}=\int dz d^2 \kappa
X^\mu_+ X^\nu_-$. These covariantly transforming superfields do not
have free-field operator product expansions, but instead obey
the relation, $\Psi_-^\alpha \Psi_+^\beta =\gamma_\mu^{\alpha\beta}
(X^\mu_+ -X^\mu_-)$, which is a ten-dimensional generalization
of the usual analyticity condition in four dimensions.$^{17,18}$

The gauge-fixed physical vertex operators for the massless states
that were introduced in reference 15 can now easily be extended to
manifestly Lorentz-covariant vertex operators. The right-moving
part of the bosonic vertex operator is simply $\eta_\mu\int dz d^2\kappa
X^\mu_+ \exp (ik\cdot X_-)$, while the right-moving part of the
fermionic vertex operator is $u_\alpha\int dz d^2\kappa
\Psi^\alpha_+\exp(ik\cdot X_-)$. Because the SO(9,1)
super-Poincar\' e generators are superconformally invariant, the
scattering amplitudes obtained by evaluating correlation functions
of these covariant vertex operators on N=2 super-Riemann surfaces
are guaranteed to be super-Poincar\'e covariant.

In order to explicitly evaluate these correlation functions, the
$X^\mu_\pm$ and $\Psi^\alpha_\pm$ superfields must be re-expressed
in terms of the original free fields, whose correlation functions
on arbitrary genus surfaces were calculated in reference 15. Although
this procedure breaks the manifest SO(9,1) invariance down to an
SU(4)xU(1) subgroup, it is straightforward to use knowledge of the
underlying larger invariance to write the resulting scattering
amplitudes in manifestly Lorentz-covariant notation.

In the conclusion of this paper, the issue of finiteness
will be briefly discussed, as well as the possibility
of generalizing the amplitude calculations to non-flat
target-space backgrounds.

\vskip 12pt
\centerline{\bf II. Free Fields of the N=2 Twistor-String}
\vskip 12 pt

The free-field action for the N=2 twistor-string corresponding to
the heterotic Green-Schwarz superstring is:$^{14}$
$$S=\int dz d\bar z d\kp d\km [{1\over 2}
(X^{+\bar l}\dzbar X^{-l}-X^{-l}\dzbar X^{+\bar l}) +
W^-\dzbar\Psi^+ -W^+\dzbar\Psi^- +\Phi^{+q}
\Phi^{-q}]\eqno(II.1)$$
where $[z,\kp,\km]$ and $[\bar z]$ are holomorphic and
anti-holomorphic coordinates on the N=(2,0) super-Riemann
surface (although this Euclidean surface is not well-defined since
$(\kpm)^*$ does not exist, independence of the right and left-moving
sectors allows heterotic superstring amplitudes to be defined
in the usual way by taking the holomorphic square-root of
non-heterotic amplitudes),
$D_{\pm}=\partial_\kpm+\half\kmp\dz$, $X^\plb$ and $X^\ml$
for $l=1$ to 4 describe the transverse degrees of freedom and
transform as $\bar 4_{+\half}$ and $4_{-\half}$ under the
remaining SU(4)xU(1) invariance, $\Dpm W^\pm$
and $\Psi^\pm$ describe the longtitudinal degrees of freedom and
transform as $1_{\pm \half}$ under SU(4)xU(1), and
$\Phi^{\pm q}|_{\kp=\km=0}$ for $q$=1 to 16 describe the heterotic
lattice and are SU(4)xU(1) scalars.
These N=(2,0) superfields are restricted by the chirality constraints:
$$\Dm X^{+\bar l}=\Dp X^{-l}=\Dm\Psi^+=\Dp\Psi^-
=\Dm\Phi^{+q}=\Dp
\Phi^{-q};\eqno(II.2)$$
the global constraint:
$$\Dp\Psi^+\Dm\Psi^--\Psi^+\dz\Psi^--\Psi^-\dz\Psi^+=\dz X^{9+0}
\hbox { for some real superfield } X^{9+0};\eqno(II.3)$$
and the super-Virasoro constraints:
$$\Dp W^+ \Dm\Psi^- -\Dm W^- \Dp\Psi^++
\Dp X^{+\bar l}\Dm X^{-l}=$$
$$\kp\km[\dzbar X^\plb \dzbar X^\ml +\dzbar X^{9-0}\dzbar X^{9+0}
+\half (\Phi^{+q}\dzbar\Phi^{-q}+\Phi^{-q}\dzbar\Phi^{+q})]=0,\eqno(II.4)$$
where $X^{9-0}\equiv {{\Dp W^+}\over {\Dp \Psi^+}}+{{\Dm W^-}\over {\Dm
\Psi^-}}$.

This free-field action is obtained by gauge-fixing the manifestly
super-Poincar\'e invariant N=(2,0) twistor-string action, which has
been shown to have the same classical degrees of freedom as the
heterotic Green-Schwarz superstring action.$^{19}$
Evidence for the quantum consistency of the free-field action
comes from the vanishing of the superconformal anomaly and from the
fact that after integrating out the $\Psi^\pm$ and $W^\pm$
superfields, one recovers the usual light-cone gauge Green-Schwarz
action, including the interaction-point operators.$^{14}$

In order to eliminate unnecessary notation, only the heterotic
superstring will be discussed in this paper although all techniques
straightforwardly generalize to the non-heterotic cases. For
example, the free-field action for the non-heterotic superstring
is obtained from equation (II.1) by extending the worldsheet to an
N=(2,2) surface and replacing the left-moving superfields, $\Phi^{\pm
q}$, with $\bar W^\pm$ and $\bar \Psi^\pm$
superfields.$^{15}$ The only subtle point is that
in the Type IIA theory, the roles of $X^{9+0}$ and
$X^{9-0}$ are reversed in the two different
sectors, i.e. the right-moving global constraint defines
$X^{9+0}$, whereas the left-moving global
constraint defines $X^{9-0}$.

As discussed in reference 15, it is convenient to solve the global
constraint of equation (II.3) by bosonizing the components of
$\Psi^\pm=\psi^\pm +\kpm \lpm$ and $\Dpm W^\pm=w^\pm+\kmp\vepm$
in the following way:
$$\lp=(\dzxp)
 e^\hp +e^{-\hm},\quad \lm=e^{-\hp}\eqno(II.5)$$
$$\wp=e^\hp (\dz \hp+\dz\hm+x^{9-0} (\dzxp))+x^{9-0} e^{-\hm},\quad
 \wm=x^{9-0} e^{-\hp},$$
where $h^\pm$ are chiral bosons with screening charge $+1$ that
satisfy $\hp (y) \hm (z) \to \log (y-z)$ as $y\to z$.

In terms of these bosonized fields, the right-handed super-Virasoro
constraints of equation (II.4) take the form:
$$T=\Gplb\Gml -\dz\hp +\dz\hm,\quad
G_-=\dz\xplb\Gml+(\hvep +\half\psi^+\dz x^{9-0}) e^{-\hp},\quad
G_+=\dz\xml\Gplb+ \eqno(II.6)$$
$$(\hvem+\half \dz x^{9-0}\psi^- )
((\dzxp)
 e^\hp +e^{-\hm})-e^\hp ((\dz h^+ +\dz h^-)\dz\psi^- +{3\over 4}
\partial^2_z \psi^-),$$
$$L=
\dz x^\plb \dz x^\ml -\half (\Gplb\dz\Gml +\Gml\dz\Gplb)-\hvep
\dz\psi^- -\hvem\dz\psi^+ +\dz \hp\dz\hm +\half(\partial^2_z \hp
+\partial^2_z \hm ),$$
where $X^\plb=x^\plb+\kp \Gplb$,
$X^\ml=x^\ml+\km \Gml$,
and $\hat\varepsilon^\pm\equiv\varepsilon^\pm -\half\dz x^{9-0} \psi^\pm
-x^{9-0}\dz \psi^\pm$.
\vskip 24pt
\centerline {\bf III. SO(9,1) Lorentz Generators}
\vskip 12pt

The simplest way to construct the SO(9,1) Lorentz generators
out of these free fields is to look for a generalization of the
operator, $m^{\mu\nu}=\int dz~ x^\mu \dz x^\nu$, that is N=(2,0)
superconformally invariant (the left-moving contribution to the
Lorentz generator, $\int d\bar z~ x^\mu \dzbar x^\nu$, is
already superconformally invariant). Since for the 15 SU(4)
generators that preserve the gauge-fixing, these generalized
operators are simply $M^{\plb ~-m}=\int dz d^2\kappa X^\plb X^\mm$,
it is natural to look for chiral and anti-chiral superfields,
$X^\mu_\pm$, that contain $x^\mu$ in their $\kpm =0$ component.

Using the free-field commutation relations, it is easy to show
that $[G_+,x^\mu_+]=[G_-,x^\mu_-]=0$, where
$$x^{9+0}_+\equiv x^{9+0} -\half\psi^+\psi^-,~~
x^{9-0}_+\equiv x^{9-0}+e^{\hp +\hm},~~
x^\plb_+\equiv x^\plb -e^\hm \psi^+ \Gplb,~~
x^\ml_+\equiv x^\ml;$$
$$x^{9+0}_-\equiv x^{9+0} +\half\psi^+\psi^-,~~
x^{9-0}_-\equiv x^{9-0},~~
x^\plb_-\equiv x^\plb,~~
x^\ml_-\equiv x^\ml-e^\hp \psi^-\Gml;\eqno(III.1)$$
and $G_\pm$ are defined in equation (II.6).
Therefore, the desired chiral and anti-chiral superfields are
$$X^\mu_+ (z^-,\km )=x^\mu_+ (z^- )+\km [G_- ,x^\mu_+ (z^- )] \quad
\hbox{ and }\quad
X^\mu_- (z^+,\kp )=x^\mu_- (z^+ )+\kp [G_+ ,x^\mu_- (z^+ )],\eqno(III.2)$$
where $z^\pm\equiv z\pm\half\kp\km$. Note that $x^\mu_+ =[G_+,x^\mu_-
\psi^+ e^\hm ]$,
$x^\mu_- =[G_-,x^\mu_+
\psi^- e^\hp ]$,
and $X^\mu_+ (y^-,\xi^- )~X^\nu _- (z^+,\kp )\to \eta^{\mu\nu}
\log (y^- - z^+ -\xi^- \kp )$ as $y\to z$.

It can now be straightforwardly checked using equations (II.6), (III.1),
and (III.2),
that the 45 operators, $$M^{\mu\nu}=\int dz d^2 \kappa X^\mu_+
X^\nu_- ~~
(=-\int dz d^2 \kappa X^\nu_+
X^\mu_- \hbox{  for  } \mu\neq\nu),\eqno(III.3)$$
are superconformally invariant, commute with each other to form
an SO(9,1) algebra, and transform $X^\mu_+$ and $X^\mu_-$ like
covariant SO(9,1) vectors.

By commuting $M^{\mu\nu}$ with the chiral and anti-chiral superfields
$\Psi^\pm$, one finds that $\Psi^+$ transforms as the $1_{+1}$
component of an SO(9,1) Weyl spinor superfield
$\Psi^\alpha_-$,
and
$\Psi^-$ transforms as the $1_{-1}$
component of an SO(9,1) Weyl spinor superfield
$\Psi^\alpha_+$, where
$$\Psi^\alpha_- (z^+,\kp)=\psi^\alpha_- (z^+)+\kp\{ G_+,\psi^\alpha_- (z^+)\},
\qquad
\Psi^\alpha_+ (z^-,\km)=\psi^\alpha_+ (z^-)+\km\{ G_-,\psi^\alpha_+ (z^-)\}
;\eqno(III.4)$$
$$\psi^+_-=\psi^+,~~
\psi^{lm}_-=e^{\hp-\hm}\psi^- \Gml\Gmm,~~
\psi^-_-={1\over {24}}
e^{2\hp-2\hm} \psi^- \dz\psi^- \hvem \epsilon_{lmnp} \Gml\Gmm
\Gamma^{-n} \Gamma^{-p},$$
$$\psi^{+l}_-=e^\hp \Gml,~~
\psi^{-\bar l}_-={1\over 6} e^{2\hp-\hm} \psi^- \hvem \epsilon_{lmnp} \Gmm
\Gamma^{-n} \Gamma^{-p};$$
$$\psi^+_+={1\over {24}}
e^{2\hm-2\hp} \psi^+ \dz\psi^+ \hvep \epsilon^{lmnp} \Gplb\Gpmb
\Gamma^{+\bar n} \Gamma^{+\bar p},~~
\psi^{lm}_+=\half e^{\hm-\hp} \psi^+ \epsilon^{lmnp}\Gamma^{+\bar n}
\Gamma{+\bar p},~~
\psi^-_+=\psi^-,$$
$$\psi^{+l}_+={1\over 6}e^{2\hm-\hp} \psi^+ \hvep \epsilon^{lmnp} \Gpmb
\Gamma^{+\bar n} \Gamma^{+\bar p},~~
\psi^{-\bar l}_+=e^\hm \Gplb;$$
and the sixteen components of the SO(9,1) Weyl spinor have been
split up into $[1_{+1},6_0,1_{-1},4_{+1},\bar 4_{-1}]$ representations
of SU(4)xU(1).

As was mentioned in the introduction, these spinor superfields
obey the identity, $$\Psi^\alpha_-\Psi^\beta_+=(X^\mu_+-X^\mu_-)
\gamma_\mu^{\alpha\beta}\eqno(III.5),$$ which can easily be checked using their
free-field expansions. This identity
is an obvious generalization of the four-dimensional analyticity
condition,$^{17,18}$
$\theta^\alpha\bar\theta^{\dot\alpha}=(x^\mu_+-x^\mu_-)
\gamma_\mu^{\alpha \dot\alpha}$, however because the ten-dimensional
identity contains 256 components, it can not be solved using purely
classical fields.

\vskip 24pt
\centerline {\bf IV. Covariant Vertex Operators}
\vskip 12pt

Using the $X^\mu_\pm$ and $\Psi^\alpha_\pm$ superfields, it
is easy to construct the superconformally invariant vertex
operators that covariantly describe the massless bosonic and
fermionic states. The massless bosonic vertex operator is
$$\eta_\mu\int d^2 z d^2\kappa X^\mu_+ e^{ik\cdot X_-}V_L~~\qquad (=-
\eta_\mu\int d^2 z d^2\kappa X^\mu_- e^{ik\cdot X_+}V_L),\eqno(IV.1)$$
where the left-moving contribution
to the vertex operator,
$V_L$, is constructed in the usual way for the heterotic
superstring$^{20}$ out of $\dzbar X^\mu_+$ and $\Phi^{\pm q}$ (note that
only the $\kpm=0$ components of these superfields contribute), and
$k^2=\eta\cdot k=0$. This vertex operator is invariant under the
gauge transformations, $\eta^\mu\to \eta^\mu+\Lambda k^\mu$, and
after fixing the gauge $\eta^{9+0}=0$ and replacing the $\int d^2 z$
integration with $c \bar c$ ghosts, it coincides with the massless
bosonic vertex operator of reference 15 in the picture with
ghost-number (1,1) and instanton-number zero (the instanton-number
of a vertex operator is defined as its eigenvalue under commutation
with the charge, $K=\half\int dz (\hvep\psi^- -\hvem\psi^++\dz\hm
-\dz\hp)$, and can be shifted by an arbitrary integer by
attaching instanton-number-changing operators to the vertex operator).

Two covariant choices for the massless fermionic vertex operator
are $$u_\alpha\int d^2 z d^2 \kappa \Psi^\alpha_- e^{ik\cdot X_+} V_L
\quad\hbox {  or  }\quad
u_\alpha\int d^2 z d^2 \kappa \Psi^\alpha_+ e^{ik\cdot X_-} V_L,\eqno(IV.2)$$
where $k^2=0$. These vertex operators are invariant under the
gauge transformations, $u_\alpha\to u_\alpha+k_\mu\Lambda^\beta\gmu$,
and after fixing the gauge $(\gamma^{9+0}u)^\alpha =0$
and replacing the $\int d^2 z$ integration with $c\bar c$ ghosts,
they coincide with the massless fermionic vertex operator of
reference 15, either in the picture with ghost-number (1,1) and
instanton-number $-\half$, or in the picture with ghost-number
(1,1) and instanton-number $+\half$.

Since the usual fermionic field, $\zeta^\alpha$, satisfies
$k_\mu \zeta^\alpha \gmu=0$ and has no gauge invariances, it is clear
that $\zeta^\alpha$ corresponds to $k^\mu u_\beta \gamma_\mu^{\alpha\beta}$.
One therefore still needs to construct the vertex operator for the
fermionic field when $k^\mu=0$. Two choices for this zero-momentum
vertex operator are
$$\zeta^\alpha \int d^2 z d^2\kappa
\gamma_{\mu ,\alpha\beta}X^\mu_+\Psi^\beta_- V_L
\quad\hbox {  or  }\quad
\zeta^\alpha \int d^2 z d^2\kappa
\gamma_{\mu ,\alpha\beta}X^\mu_-\Psi^\beta_+ V_L,\eqno(IV.3)$$
out of which can be extracted two choices for the covariant
spacetime supersymmetry generators,
$$S_{-,\alpha}= {2\over 5}\int dz d^2\kappa
\gamma_{\mu ,\alpha\beta}X^\mu_+\Psi^\beta_-
\quad\hbox {  or  }\quad
S_{+,\alpha}= {2\over 5}\int dz d^2\kappa
\gamma_{\mu ,\alpha\beta}X^\mu_-\Psi^\beta_+ ,\eqno(IV.4)$$
which satisfy $\{ S_{-,\alpha},S_{+,\beta}\} =2\int dz \dz x_\mu\gmu$. It is
easily verified that commutation with
$S_{\pm,\alpha}$ takes the fermionic vertex
operator of instanton-number $\mp\half$ and spinor polarization $u_\beta$
into the bosonic vertex operator of vector polarization $\eta^\mu=
(\gamma^\mu \gamma^\nu u)_\alpha k_\nu$,
and takes the bosonic vertex operator of vector polarization $\eta^\mu$
into the fermionic vertex operator of instanton-number $\pm\half$
and spinor-polarization $u_\beta=\gmu \eta_\mu$.

It is interesting to note that the integrands of the spacetime
supersymmetry generators, $$W_{-,\alpha}=\gamma_{\mu,\alpha\beta}
X^\mu_+ \Psi^\beta_- \quad\hbox{  and  }\quad
W_{+,\alpha}=\gamma_{\mu,\alpha\beta}
X^\mu_- \Psi^\beta_+,\eqno(IV.5)$$ satisfy the free-field behavior
$$D_+ W_-(y^-,\xi^-)~\Psi_+(z^-,\km)\to (\xi^--\km)/(y^- -z^-)\hbox{  and  }
D_- W_+(y^+,\xi^+)~\Psi_-(z^+,\kp)\to (\xi^+-\kp)/(y^+ -z^+)$$
as $y\to z$, and therefore provide an N=(2,0) generalization of the
twistor equation,$^{21,22}$
$w_\alpha =
\gamma_{\mu,\alpha\beta}x^\mu \lambda^\beta$.

\vskip 24pt
\centerline {\bf V. Lorentz-Covariant Scattering Amplitudes}
\vskip 12pt

Scattering amplitudes can now be calculated by evaluating correlation
functions of these covariant vertex operators on N=2 super-Riemann surfaces
and integrating over the global super-moduli of the surfaces. By
choosing half of the fermionic vertex operators to have
instanton-number $+\half$ and the other half to have instanton-number
$-\half$, the N=2 super-Riemann surfaces can be restricted to those
with vanishing instanton-number (in order to get a non-zero
correlation function, the instanton-number of the surface must equal
the sum of the instanton-numbers of the vertex operators). The
choice of which fermionic vetex operators have $+\half$ or $-\half$
instanton-number does not affect the scattering amplitude since
different choices correspond to different locations for the
picture-changing and instanton-number-changing operators (see
reference 15 for more details). Similarly, the choice of using
$S_{+,\alpha}$ or $S_{-,\alpha}$ does not affect the scattering
amplitude as long as the instanton-number of the surface is set
equal to the total instanton-number of the vertex operators
plus the total instanton-number of the spacetime supersymmetry
generators.

Since $M^{\mu\nu}$ and $S_{\pm,\alpha}$ of equations (III.3) and (IV.4)
are superconformally invariant and covariantly transform the
vertex operators, these scattering amplitudes are proven to be
SO(9,1) super-Poincar\'e covariant.

At the present time, the only known way to evaluate correlation
functions of the covariant vertex operators is to re-express the
vertex operators in terms of the original set of free fields,
and to use the free-field correlation functions that were
evaluated on arbitrary genus surfaces in reference 15. Although
this procedure breaks the manifest Lorentz covariance down to
an SU(4)xU(1) subgroup, knowledge of the underlying larger invariance
can be used to express the resulting scattering amplitudes in
manifestly Lorentz-covariant notation.

The first step in covariantizing the calculation is to introduce
an SO(9,1) pure spinor$^{23}$ and its complex conjugate, $v^\alpha$
and $\vba$, and a real SO(9,1) light-like vector, $\m$, which
satisfy
$$\va\gmu\vb =\vba\gmu\vbb=\m m_\mu=0\quad\hbox{  and  }\quad
\va\gmu\vbb m_\mu=1.\eqno(V.1)$$
With the help of these projection operators, the free fields
$\Gplb$ and $\Gml$ can be treated as four complex components of
SO(9,1) vectors $\Gamma^\mu_-$ and $\Gamma^\mu_+$ (all other
components of the vectors will be projected out), while the
free fields $\psi^\pm$, $\hvepm$, $h^\pm$, as well as all
ghost fields (see reference 15 for details on the ghost fields),
can be treated as SO(9,1) scalars. For example, the $\kpm=0$ component
of the covariant superfields $X^\mu_-$ and $\Psi^\alpha_-$ can
be written as
$$x^\mu_-=x^\mu +\half\psi^+\psi^- (v\gamma^\mu \bar v)-e^\hp\psi^-
m^\nu \Gamma_+^\rho
(v\gamma^\mu \gamma_\nu \gamma_\rho \bar v),\eqno(V.2)$$
$$\psi^\alpha_-=\psi^+ \vba + e^\hp
\m \Gamma^\nu_+ (\gamma_\mu \gamma_\nu \bar v)^\alpha
+{1\over 4} e^{\hp-\hm}\psi^- m^\rho\Gamma^\sigma_+
\Gamma^\tau_+  (\bar v\gamma_\mu\gamma_\nu\gamma_\rho\gamma_\sigma
\gamma_\tau \bar v)(\gamma^\mu\gamma^\nu v)^\alpha+$$
$${1\over 6} e^{2\hp-\hm}\psi^-\hvem \m \Gamma^\nu_+ \Gamma^\rho_+
\Gamma^\sigma_+
(\gamma_\mu\gamma_\nu\gamma_\rho\gamma_\sigma \bar v)^\alpha +
{1\over {24}}e^{2\hp-2\hm}\psi^-\dz\psi^-\hvem \m \Gamma^\nu_+ \Gamma^\rho_+
\Gamma^\sigma_+ \Gamma^\tau_+
(\bar v\gamma_\mu\gamma_\nu\gamma_\rho\gamma_\sigma \gamma_\tau \bar v)\va,$$
and the term $\dz x^{\ml} \Gplb$ in $G_+$ of equation (III.6) can be
written as $\dz x^\mu m^\nu \Gamma^\rho_- (\bar v \gamma_\mu\gamma_\nu
\gamma_\rho  v)$.

After evaluating correlation functions of the free fields as in
reference 15, one is left with an expression for the scattering
amplitude that is manifestly Lorentz covariant, but which is a
polynomial in the three projection operators, $\va$, $\vba$, and $\m$.
However, the knowledge that the scattering amplitude is Lorentz
covariant even without transforming the projection operators implies
that the projection operators only occur in Lorentz-invariant
combinations. Therefore, replacing all monomials of projection
operators by their Lorentz-invariant component does not affect
the scattering amplitude, since all non-Lorentz-invariant
components must cancel out. So covariantization of the Green-Schwarz
superstring amplitudes has been reduced to the problem of extracting
the Lorentz-invariant component of the monomial
$v^{\alpha_1} ... v^{\alpha_L} \bar v^{\beta_1} ... \bar v^{\beta_M}
m^{\mu_1} ... m^{\mu_N}$, where $\va$, $\vba$, and $\m$ satisfy
equation (V.1). Note that this covariantization procedure requires that
the amplitude does not contain negative powers of the projection
operators, and therefore is not useful in the light-cone or
semi-light-cone Green-Schwarz formalisms where $(k^{9+0})^{-1}$
factors are present.

Since under a U(1) rotation, $\va\to e^{i\phi}\va$, $\vba\to e^{-i\phi}
\vba$, $\m\to\m$; while under an SO(1,1) boost,
$\va\to e^{\phi}\va$, $\vba\to e^{\phi}
\vba$, $\m\to e^{-2\phi}\m$; the Lorentz-invariant component of
a monomial is zero unless there are an equal number of $\va$'s,
$\vba$'s and $\m$'s. The Lorentz-invariant component,
$P_N^{\alpha_1 ... \alpha_N \beta_1 ... \beta_N
\mu_1 ... \mu_N}$,
of all such contributing monomials,
$v^{\alpha_1} ... v^{\alpha_N} \bar v^{\beta_1} ... \bar v^{\beta_N}
m^{\mu_1} ... m^{\mu_N}$, can be obtained from the formula:
$$P_N^{\alpha_1 ... \alpha_N \beta_1 ... \beta_N
\mu_1 ... \mu_N}=
c_N\sum_{symmetrized}
\eqno(V.3)$$
$$[(N!)^{-1}
\gamma^{\mu_1,\alpha_1 \beta_1} ...
\gamma^{\mu_N,\alpha_N \beta_N} + {{(2(N-2)!)^{-3}}\over {N+2}}
\gamma_\nu^{\alpha_1 \alpha_2}
\gamma_\rho^{\beta_1 \beta_2}
(\eta^{\nu\rho}\eta^{\mu_1 \mu_2}-2
\eta^{\nu \mu_1}\eta^{\rho \mu_2})
P_{N-2}^{\alpha_3 ... \alpha_N \beta_3 ... \beta_N
\mu_3 ... \mu_N}],$$
where $\sum$ means to symmetrize independently in the
$\alpha$, $\beta$, and $\mu$ indices to give $(N!)^3$ terms, $P_0 =1$,
$P_1^{\alpha\beta\mu}={1 \over {160}}\gamma^{\mu,\alpha\beta}$,
and $c_N$ is fixed by the requirement that
$
\gamma_{\mu_1,\alpha_1 \beta_1} ...
\gamma_{\mu_N,\alpha_N \beta_N}
P_N^{\alpha_1 ... \alpha_N \beta_1 ... \beta_N
\mu_1 ... \mu_N}=1$. It is straightforward to check that
$P_N^{\alpha_1 ... \alpha_N \beta_1 ... \beta_N
\mu_1 ... \mu_N}$ is the unique Lorentz-invariant tensor
which is symmetric in the
$\alpha$, $\beta$, and $\mu$ indices,  which vanishes when contracted
with $\gamma^\sigma_{\alpha_1 \alpha_2}$,
$\gamma^\sigma_{\beta_1 \beta_2}$, or $\eta_{\mu_1 \mu_2}$,
and which is normalized to one when all indices are contracted
with $\gamma$-matrices. So after replacing all contributing monomials
by $P_N$ in the expression for the scattering amplitude that is
calculated using the free-field description of the covariant vertex
operators, the Green-Schwarz superstring amplitudes are written in
manifestly Lorentz-covariant notation.
\vskip 24pt
\centerline {\bf VI. Conclusion}
\vskip 12pt

After constructing SO(9,1) super-Poincar\'e generators and
manifestly covariant vertex operators, it was proved in this paper
that the earlier calculated multiloop Green-Schwarz superstring
scattering amplitudes are super-Poincar\'e invariant, and it
was shown how to write these amplitudes in manifestly Lorentz-covariant
form.

One possible use for these amplitudes is to analyze their finiteness
properties. Using the Neveu-Schwarz-Ramond method, it is difficult
to show that the dilaton tadpole diagram vanishes,$^5$ while using the
light-cone Green-Schwarz method, it is difficult to show that
the contact terms precisely cancel the divergences when two or
more interaction points coalesce.$^9$ Since neither of these difficulties
are present using the twistor-string method, it should be
possible to check for finiteness by explicitly looking for divergences
in the scattering amplitude expressions.

In the construction of the super-Poincar\'e generators and covariant
vertex operators, the N=2 chiral and anti-chiral superfields
$X^\mu_\pm$ and $\Psi^\alpha_\pm$ played a prominent role. It is
tempting to suggest that in a non-flat target-space background, these
N=2 superfields provide the phase space for the ten-dimensional
supergravity and super-Yang-Mills fields.$^{24}$
In flat space, the
superfields obey the relations,
$\Psi^\alpha_-\Psi^\beta_+=(X^\mu_+-X^\mu_-)
\gamma_\mu^{\alpha\beta}$ from
equation (III.5), and $W_{+,\alpha}=\gamma_{\mu,\alpha\beta}
X^\mu_- \Psi^\beta_+$ from equation
(IV.5), where $X^\mu_+(y^-,\xi^-) X^\nu_- (z^+,\kp)
\to \log (y^- -z^+ -\xi^-\kp)$ and
$D_- W_+(y^+,\xi^+)\Psi_-(z^+,\kp)\to (\xi^+-\kp)/(y^+ -z^+).$
Perhaps in a curved target-space background, these superfield
relations are modified in the same way that the four-dimensional
analyticity condition,
$\theta^\alpha\bar\theta^{\dot\alpha}=(x^\mu_+-x^\mu_-)
\gamma_\mu^{\alpha \dot\alpha}$, is modified to $H(x,
\theta,\bar\theta)^{\alpha
\dot\alpha}=
(x^\mu_+-x^\mu_-)
\gamma_\mu^{\alpha \dot\alpha}$ in the presence of supergravity.$^{25}$

\vskip 24pt
\centerline {\bf Acknowledgements}
\vskip 12pt
I would like to thank M. Green, P. Howe, S. Mandelstam,
W. Siegel, J. Taylor, and
P. Townsend
for useful discussions, and the SERC for its financial support.

\vskip 24pt

\centerline{\bf References}
\vskip 12pt

\item{(1)} Friedan,D., Martinec,E., and Shenker,S., Nucl.Phys.B271
(1986), p.93.

\item{(2)} Verlinde,E. and Verlinde,H., Phys.Lett.B192 (1987), p.95.

\item{(3)} Aoki,K., D'Hoker,E., and Phong,D.H., Nucl.Phys.B342 (1990),
p.149.

\item{(4)} Atick,J. and Sen,A., Nucl.Phys.B296 (1988), p.157.

\item{(5)} Mandelstam,S., ``The n-loop Amplitude: Explicit Formulas,
Finiteness, and Absence of Ambiguities'', preprint UCB-PTH-91/53,
October 1991.

\item{(6)} Iengo,R. and Zhu,C.-J., Phys.Lett.B212 (1988), p.313.

\item{(7)} Green,M.B. and Schwarz,J.H., Nucl.Phys.B243 (1984), p.475.

\item{(8)} Mandelstam,S., Prog.Theor.Phys.Suppl.86 (1986), p.163.

\item{(9)} Restuccia,A. and Taylor,J.G., Phys.Rep.174 (1989), p.283.

\item{(10)} Greensite,J. and Klinkhamer,F.R., Nucl.Phys.B291 (1987), p.557.

\item{(11)} Mandelstam,S., private communication.

\item{(12)} Carlip,S., Nucl.Phys.B284 (1987), p.365.

\item{(13)} Gilbert,G. and Johnston,D., Phys.Lett.B205 (1988), p.273.

\item{(14)} Berkovits,N., Nucl.Phys.B379 (1992), p.96., hep-th bulletin
board 9201004.

\item{(15)} Berkovits,N., ``Calculation of Greeen-Schwarz Superstring
Amplitudes using the N=2 Twistor-String Formalism'', SUNY at Stonybrook
preprint ITP-SB-92-42, August 1992, hep-th bulletin board 9208035.

\item{(16)} Martinec,E., Phys.Lett.B171 (1986), p.189.

\item{(17)} Delduc,F. and Sokatchev,E., Class.Quant.Grav. 9 (1992), p.361.

\item{(18)} Ivanov,E.A. and Kapustnikov,A.A., Phys.Lett.B267 (1991), p.175.

\item{(19)} Tonin,M., Phys.Lett.B266 (1991), p.312.

\item{(20)} Gross,D.J., Harvey,J.A., Martinec,E., and Rohm,R.,
Nucl.Phys.B256 (1985), p.253.

\item{(21)} Penrose,R. and MacCallum,M.A.H., Phys.Rep.6C (1972), p.241.

\item{(22)} Sorokin,D.P., Tkach,V.I., Volkov,D.V., and Zheltukhin,A.A.,
Phys.Lett.B216 (1989), p.302.

\item{(23)} Howe,P.S., Phys.Lett.B258 (1991), p.141.

\item{(24)} Gates,S.J.Jr. and Nishino,H., Class.Quant.Grav. 3 (1986), p.39.

\item{(25)} Ogievetsky,V.I. and Sokatchev,E.S., Phys.Lett.79 (1978), p.222.

\end